\documentclass[pdflatex,sn-basic,Numbered,hidelinks]{sn-jnl}
\usepackage{graphicx}%
\usepackage{multirow}%
\usepackage{amsmath,amssymb,amsfonts}%
\usepackage{amsthm}%
\usepackage{mathrsfs}%
\usepackage[title]{appendix}%
\usepackage{xcolor}%
\usepackage{textcomp}%
\usepackage{manyfoot}%
\usepackage{booktabs}%
\usepackage{algorithm}%
\usepackage{algorithmicx}%
\usepackage{algpseudocode}%
\usepackage{listings}%
\usepackage{graphicx}
\usepackage{makecell}
\usepackage{subfig}
\usepackage{svg}

\theoremstyle{thmstyleone}%
\theoremstyle{thmstyletwo}%

\theoremstyle{thmstylethree}%

\begin{document}

\title[Article Title]{An Explainable and Scalable Hierarchical Tensor Framework for Integrating Multi-Omics and Digital Pathomics}


%
\author[1]{\fnm{Ahmad} \sur{Hussein}}\email{ahmad.hussein@uts.edu.au}

\author[2,3]{\fnm{Ali} \sur{Anaissi}}\email{ali.anaissi@uts.edu.au}

\author[1]{\fnm{Mukesh} \sur{Prasad}}\email{mukesh.prasad@uts.edu.au}

\author*[1]{\fnm{Ali} \sur{Braytee}}\email{ali.braytee@uts.edu.au}

\affil*[1]{\orgdiv{School of Computer Science}, \orgname{University of Technology Sydney}, \orgaddress{\city{Sydney}, \state{NSW}, \country{Australia}}}

\affil[2]{\orgdiv{TD School}, \orgname{University of Technology Sydney}, \orgaddress{\city{Sydney}, \state{NSW}, \country{Australia}}}

\affil[3]{\orgdiv{School of Computer Science}, \orgname{The University of Sydney}, \orgaddress{\city{Sydney}, \state{NSW}, \country{Australia}}}



\abstract{ \textbf{Background:} Precision oncology increasingly relies on integrating heterogeneous data spanning the biological axis from genomics and transcriptomics to proteomics and digital pathology. However, many current fusion methods prioritize predictive performance while offering limited support for structured, biologically grounded explainability, which can reduce robustness and scalability when applied to large clinical cohorts with heterogeneous and incomplete modality availability.

\textbf{Methods:} We propose a Hierarchical Interpretable Tensor Fusion, \textit{HIT-Fuse}, for integrating six modalities: somatic copy number variation, DNA methylation, RNA sequencing, microRNA expression, protein abundance (RPPA), and vision transformer-extracted pathomics features across four TCGA cohorts: glioma, lung adenocarcinoma, breast cancer, and kidney cancer. High-dimensional inputs are first compressed into latent representations using modality-specific autoencoders, followed by a two-stage CP tensor decomposition that preserves biologically aligned structure. To handle missing modalities, we reconstruct latent representations for patients with available clinical data by leveraging structured clinical covariates together with observed omics profiles, and retrain downstream models on the expanded cohort. Global and local explainable artificial intelligence (xAI) attributions are computed through a mathematical traceback mechanism.

\textbf{Results:} Across glioma, breast, and kidney cancer cohorts, HIT-Fuse yields well-separated latent subgroups with significant survival stratification in both complete and reconstructed settings. Cohort expansion produces cohort-dependent performance gains, including a \textbf{56.4\% relative improvement in weighted F1-score for glioma}. Cross-domain biological validation integrating DepMap CRISPR dependency screens, miRTarBase interactions, and RPPA annotations confirms functional support for decoded drivers across four cancer cohorts. Essential gene rates range from 6.2\% (glioma) to 7.6\% (lung adenocarcinoma), with representative confirmed drivers including \textit{ANAPC11} (glioma), \textit{SNRPD1} (breast), \textit{PLK1} (lung), and \textit{RPL14} (kidney). Additionally, 71--92\% of decoded miRNAs show regulatory support in miRTarBase, and 77--84\% of selected genes are covered in DepMap.

\textbf{Conclusions:} By combining latent-space fusion, biologically structured tensor decomposition, and attribution-based explainability, \textit{HIT-Fuse} provides a scalable and interpretable framework for multi-omics and digital pathology integration, enabling robust cancer prognosis modeling at cohort scale.}


\maketitle

\section{Introduction}  
Precision oncology increasingly depends on the integration of heterogeneous biological data spanning multiple molecular layers and tissue-level phenotypes. Advances in high-throughput sequencing enable comprehensive profiling of genomic alterations, DNA methylation, transcriptomic activity, proteomic abundance, and microRNA regulation at population scale \cite{Hasin2017}. In parallel, computational pathology and self-supervised vision models have enabled scalable representation learning from whole-slide histopathology images, extracting high-dimensional morphological embeddings from gigapixel tissue sections \cite{Lu2021}. Together, these modalities provide a systems-level characterization of tumor biology across the central dogma, from DNA to RNA to protein and tissue phenotype. However, integrating these heterogeneous data streams into scalable, interpretable, and biologically grounded predictive models remains a central challenge. Many multi-omics integration frameworks prioritize predictive performance through feature concatenation, kernel-based fusion, or deep neural architectures that treat modalities as flat feature spaces \cite{Ritchie2015,Argelaguet2018}. While such approaches can achieve strong classification or survival performance, they often underrepresent the hierarchical organization of biological systems, limiting mechanistic interpretability and clinical transparency \cite{Holzinger2017,Lipton2018,Kitano2002}. Tensor-based representations offer a principled alternative for modeling structured multi-way relationships in high-dimensional biomedical data. By preserving interactions between patients, modalities, and latent biological factors, tensor decompositions extend beyond conventional matrix-based fusion \cite{Kolda2009}. Prior work has applied CANDECOMP/PARAFAC (CP) and Tucker decompositions to multi-omics feature extraction, subtype discovery, and survival modeling \cite{Yang2021IEEE}. More recently, scalable tensor factorization strategies have been developed for large biomedical datasets, highlighting their suitability for big data contexts \cite{Smith2017}. Nevertheless, most tensor-driven approaches operate at a single, flat fusion level, potentially diluting biologically meaningful structure across molecular layers. Concurrently, multimodal learning frameworks integrating digital pathology with molecular profiles have demonstrated improved prognostic modeling and molecular subtype prediction \cite{Cheerla2019,Chen2022}. Vision transformer and self-supervised learning approaches, including DINO-based architectures, enable robust whole-slide representation learning that can be fused with omics data \cite{Lu2021}. However, image features are typically incorporated as auxiliary embeddings or fused at late stages, limiting the ability to uncover biologically aligned cross-modal interactions. A further practical challenge arises from incomplete modality availability in real-world cohorts. Large-scale resources such as The Cancer Genome Atlas (TCGA) provide rich but heterogeneous multi-omics datasets with pervasive missingness across molecular layers \cite{tcga2013pancancer}. Restricting analysis to the strict intersection of complete cases substantially reduces cohort size and statistical power. While probabilistic multi-view and representation learning approaches have explored cross-modal alignment and latent completion \cite{Argelaguet2018,Li2022}, few frameworks integrate biologically structured fusion, modality completion, and external biological validation within a unified and interpretable pipeline.

\begin{figure}[!htbp]
    \centering
    \includegraphics[width=\textwidth]{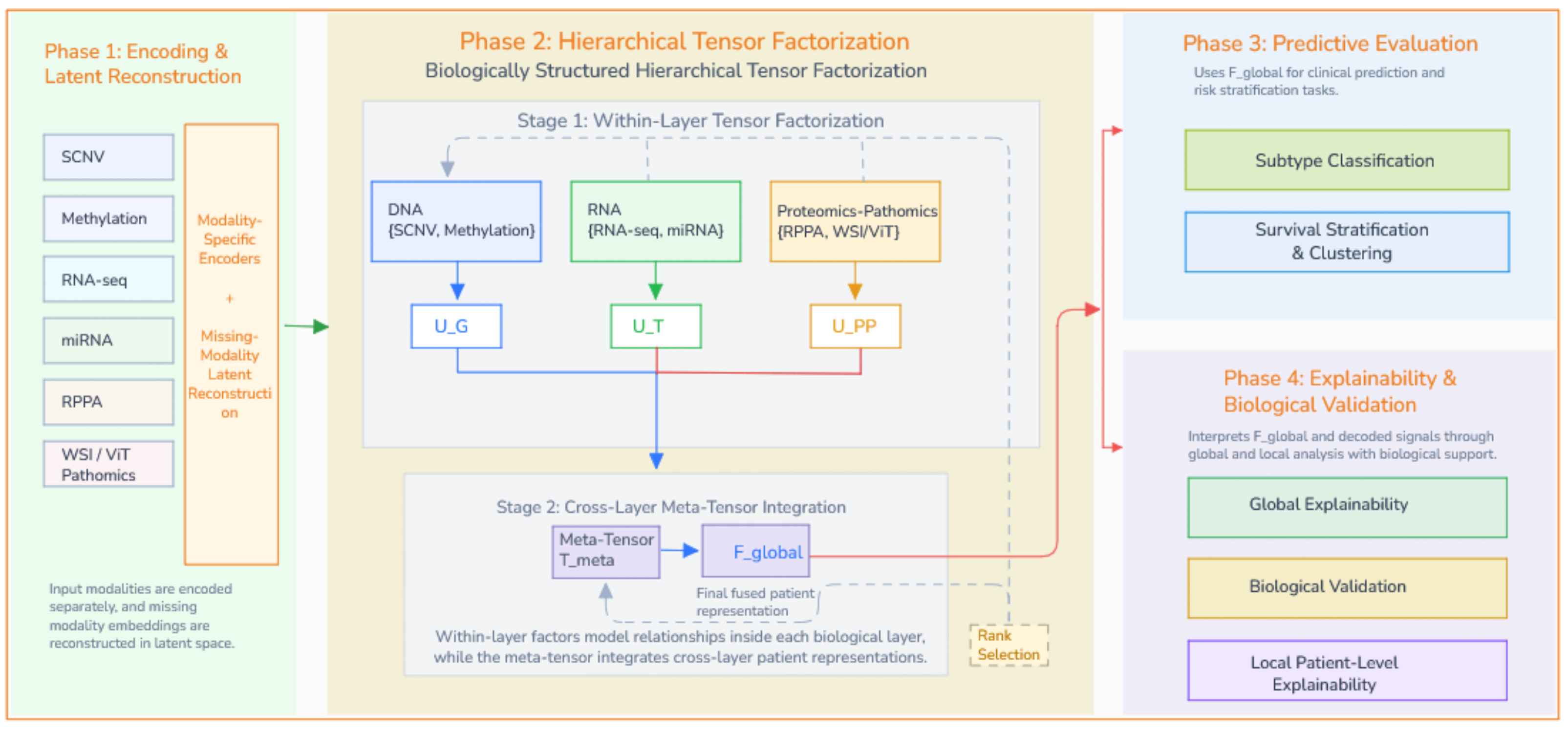}
    \caption{Overview of our proposed Hierarchical Multimodal Tensor Fusion framework (HIT-Fuse). The framework consists of four phases: modality-specific encoding and missing-modality latent reconstruction; biologically structured hierarchical tensor factorization; predictive evaluation using $F_{\text{global}}$ for subtype classification and clustering-based survival analysis; and structured explainability through global biological validation and local patient-level interpretation.}
    \label{Framework-HIT}
\end{figure}

To address these limitations, we propose an explainable and scalable hierarchical tensor framework, termed HIT-Fuse (Figure~\ref{Framework-HIT}), for integrating multi-omics and digital pathology data. The framework learns structured patient representations that respect biological organization across molecular layers while supporting interpretable subtype prediction and survival modeling. To accommodate incomplete clinical cohorts, HIT-Fuse incorporates a modality completion mechanism that enables learning beyond strict intersection subsets. We evaluate the framework across four independent TCGA cancer cohorts: glioma, lung adenocarcinoma, breast cancer, and kidney cancer.

\paragraph{Contributions.} We summarize the main contributions of this work as follows: \begin{itemize} 
\item We propose a biologically structured two-stage tensor fusion framework that integrates multi-omics and digital pathology while preserving central-dogma relationships across molecular layers, enabling interpretable patient representations at cohort scale. \item We introduce a latent completion framework that reconstructs missing modalities using observed omics and clinical covariates, enabling analysis beyond strict intersection cohorts. \item We establish an end-to-end biological validation workflow that links hierarchical tensor explainability to external functional evidence, validating discovered drivers using DepMap CRISPR dependencies and pathway enrichment across multiple cancer cohorts. \item We develop a hierarchical explainability framework that delivers both global cohort-level drivers and patient-specific molecular attributions by tracing tensor meta-factors back to raw biological features. \end{itemize}

\section{Related Work}
\label{sec:related_work}

Deep learning has become a dominant paradigm for multi-omics integration due to its ability to capture nonlinear relationships in high-dimensional molecular data. Chaudhary et al.~\cite{chaudhary2018multiomics} demonstrated that stacked autoencoders can learn modality-specific latent representations from RNA-seq, miRNA, and DNA methylation data for survival prediction in hepatocellular carcinoma. Similarly, DeepProg combined autoencoder-based embedding with classical machine learning to achieve cross-cohort prognostic modeling across TCGA datasets~\cite{poirion2021deepprog}. While these frameworks achieve strong predictive performance, they typically rely on flat latent concatenation, treating modalities as independent feature blocks and overlooking hierarchical biological organization across molecular layers. 
Probabilistic multi-view factor models have sought to improve interpretability by explicitly modeling shared and modality-specific variation. MOFA+ introduced a scalable Bayesian group factor analysis framework capable of learning structured latent representations across heterogeneous omics layers~\cite{argelaguet2020mofaplus}. Although such models improve interpretability relative to purely deep architectures, they remain fundamentally matrix-based and do not explicitly model higher-order modality interactions or hierarchical biological flow.

Tensor decomposition provides a principled alternative for modeling multi-way biomedical data. Foundational work by Kolda and Bader~\cite{kolda2009tensor} established CANDECOMP/PARAFAC (CP) and Tucker decompositions as scalable tools for structured high-dimensional analysis. Subsequent applications demonstrated the utility of tensor methods for cancer subtype discovery and survival modeling from multi-omics cohorts~\cite{yang2021tensor}. A closely related approach by ~\cite{Braytee2024} combined modality-specific autoencoders with single-stage CP tensor fusion for cancer risk stratification. While this framework showed the benefit of integrating deep representation learning with tensor analysis, it performs fusion at a single flat level and does not explicitly encode biological hierarchy, incorporate digital pathology, or provide structured end-to-end biological validation.

In parallel, multimodal learning frameworks integrating digital pathology with molecular profiles have demonstrated improved diagnostic and prognostic performance. Attention-based co-learning models such as MCAT~\cite{chen2021mcat} and transformer-based multimodal survival frameworks such as PORPOISE~\cite{porpoise2022} enable joint modeling of whole-slide images and genomic data. Self-supervised vision transformers have further strengthened histopathology representation learning~\cite{Lu2021}. However, most existing multimodal systems employ late fusion or flat concatenation strategies and treat image features as auxiliary modalities rather than embedding them within a biologically structured fusion hierarchy.

A persistent challenge in large-scale cancer cohorts is incomplete modality availability. Public resources such as TCGA provide rich but heterogeneous datasets with pervasive missingness across molecular layers. Although recent multi-view representation learning approaches have explored cross-modal alignment and completion~\cite{li2022multiview}, most do not integrate structured biological fusion with external functional validation.

Finally, explainable artificial intelligence methods have been increasingly adopted to enhance transparency in biomedical prediction models. Feature attribution techniques such as Integrated Gradients~\cite{sundararajan2017ig} and SHAP~\cite{lundberg2017shap} provide post-hoc importance scores for molecular features. However, these techniques are typically applied to flat architectures and do not inherently support hierarchical tensor-based fusion or closed-loop biological validation.

Overall, prior work has demonstrated the effectiveness of deep representation learning, tensor decomposition, multimodal pathology integration, and explainable AI in isolation. However, existing frameworks typically combine only a subset of these components. In particular, most tensor-driven and deep multi-omics approaches rely on single-stage fusion and provide limited linkage between model explanations and experimentally grounded biological evidence. These limitations motivate the development of a unified hierarchical tensor framework that integrates biologically aligned multi-modal fusion, modality completion, and end-to-end biological validation within a scalable and interpretable pipeline.

\section{Materials and Methods}
\label{sec:methods}

This section presents a scalable and biologically grounded framework for integrating heterogeneous multi-omics and histopathology data using hierarchical tensor fusion with end-to-end global and local explainability. Throughout, we distinguish between two evaluation cohorts: a \emph{real-only intersection cohort} (E1), consisting of patients with complete observations across all modalities, and a \emph{completed all-patient cohort} (E2), consisting of the full clinical population augmented via latent-space super-imputation. 


\subsection{Phase 1: Encoding and Latent Reconstruction}
\label{subsec:phase1}

\subsubsection{Modality Encoding}

The first stage of the framework learns compact, modality-specific latent representations independently for each omics and imaging view. This design isolates within-modality structure from cross-modality dependencies and ensures that subsequent completion and fusion operate in a unified, low-dimensional latent space rather than in the original high-dimensional feature domains.

For a given cancer cohort $c$ and modality $m$, let the raw feature matrix be
\begin{equation}
X^{(c,m)} \in \mathbb{R}^{|\mathcal{P}_{c,m}| \times p_m},
\end{equation}
where $p_m$ denotes the dimensionality of modality $m$ (e.g., gene-level features for RNA-seq, probe-level features for methylation, or pathomics features for imaging).

\paragraph{Preprocessing.}
Missing values within each modality are imputed using feature-wise medians computed over $\mathcal{P}_{c,m}$. Features are then linearly scaled to the interval $[-1,1]$ to match the output range of the decoder.

\paragraph{Latent dimensionality.}
The latent size $d_m$ is selected as a function of the input dimensionality $p_m$:
\begin{equation}
d_m =
\begin{cases}
512, & p_m > 2000, \\
256, & 500 < p_m \leq 2000, \\
64, & p_m \leq 500.
\end{cases}
\label{eq:latent_rule}
\end{equation}

\paragraph{Encoder--decoder mapping.}
Each modality is modeled using a feed-forward autoencoder with an encoder $f_{\theta_m}: \mathbb{R}^{p_m} \rightarrow \mathbb{R}^{d_m}$ and a decoder $g_{\psi_m}: \mathbb{R}^{d_m} \rightarrow \mathbb{R}^{p_m}$. For patient $i$, the latent representation is
\begin{equation}
z_i^{(c,m)} = f_{\theta_m}\big(x_i^{(c,m)}\big).
\label{eq:latent_extract}
\end{equation}

\paragraph{Training objective.}
Autoencoders are trained to minimize the mean squared reconstruction error over observed patients:
\begin{equation}
\min_{\theta_m,\psi_m} \; 
\frac{1}{|\mathcal{P}_{c,m}|} 
\sum_{i \in \mathcal{P}_{c,m}} 
\left\| x_i^{(c,m)} - g_{\psi_m}\big(f_{\theta_m}(x_i^{(c,m)})\big) \right\|_2^2.
\label{eq:ae_loss}
\end{equation}

The trained encoders produce modality-specific latent matrices
\(
Z^{(c,m)} \in \mathbb{R}^{|\mathcal{P}_{c,m}| \times d_m},
\)
which serve as the basis for cohort-wide completion and hierarchical fusion.


\subsubsection{Cross-Modal Latent Super-Imputation.}
\label{subsubsec:super_imputation}

A central challenge in multi-omics integration is pervasive and structured missingness, which severely limits analysis to the small subset of patients with complete observations across all modalities. To overcome this constraint, we introduce a latent-space \emph{super-imputation} framework that predicts missing modality representations from observed modalities and clinical context, enabling construction of a completed cohort-wide latent collection (E2).

\paragraph{Latent and mask construction.}
For each modality $m$ and cohort $c$, we define a cohort-wide latent matrix and an availability mask:
\begin{equation}
L^{(c,m)} \in \mathbb{R}^{|\mathcal{P}_c| \times d_m}, \quad
M^{(c,m)} \in \{0,1\}^{|\mathcal{P}_c|}.
\label{eq:latent_mask}
\end{equation}

These are populated as
\begin{equation}
M_i^{(c,m)} = \mathbb{I}[i \in \mathcal{P}_{c,m}], \quad
L_{i:}^{(c,m)} =
\begin{cases}
z_i^{(c,m)}, & M_i^{(c,m)} = 1, \\
\mathbf{0}, & M_i^{(c,m)} = 0.
\end{cases}
\label{eq:latent_fill}
\end{equation}

\paragraph{Clinical conditioning.}
Let $C^{(c)} \in \mathbb{R}^{|\mathcal{P}_c| \times q}$ denote the standardized clinical feature matrix. For patient $i$, $c_i^{(c)}$ represents the corresponding clinical vector.

\paragraph{Generator input.}
For a target modality $t$, the generator input for patient $i$ is constructed by concatenating clinical context, all modality latents, and availability indicators:
\begin{equation}
x_i^{(c,t)} = \Big[ c_i^{(c)}, \, L_{i:}^{(c,1)}, M_i^{(c,1)}, \ldots, L_{i:}^{(c,K)}, M_i^{(c,K)} \Big].
\label{eq:gen_input}
\end{equation}

\paragraph{Leakage prevention.}
During training of the generator for modality $t$, the target latent and mask are forcibly hidden:
\begin{equation}
L_{i:}^{(c,t)} \leftarrow \mathbf{0}, \quad M_i^{(c,t)} \leftarrow 0, \quad \forall i \in \mathcal{P}_c.
\label{eq:mask_target}
\end{equation}

\paragraph{Supervised training set.}
Only patients with observed target modality contribute supervision:
\begin{equation}
\mathcal{T}_{c,t} = \{ i \in \mathcal{P}_c \mid M_i^{(c,t)} = 1 \}.
\label{eq:train_set}
\end{equation}

\paragraph{Generator model and loss.}
A dedicated generator $G_{\phi_t}$ is trained for each modality $t$:
\begin{equation}
\hat{y}_i^{(c,t)} = G_{\phi_t}\big(x_i^{(c,t)}\big),
\label{eq:gen_forward}
\end{equation}
by minimizing
\begin{equation}
\min_{\phi_t} \; 
\frac{1}{|\mathcal{T}_{c,t}|} 
\sum_{i \in \mathcal{T}_{c,t}} 
\left\| \hat{y}_i^{(c,t)} - z_i^{(c,t)} \right\|_2^2.
\label{eq:gen_loss}
\end{equation}

\paragraph{Completion merge.}
The final completed latent for modality $t$ is
\begin{equation}
\tilde{z}_i^{(c,t)} =
\begin{cases}
z_i^{(c,t)}, & M_i^{(c,t)} = 1, \\
\hat{y}_i^{(c,t)}, & M_i^{(c,t)} = 0.
\end{cases}
\label{eq:final_latent}
\end{equation}

This procedure yields a complete latent representation for every patient and modality, forming the E2 cohort used for hierarchical tensor fusion.


\subsubsection{Cohort Construction and Latent Harmonization}
\label{subsubsec:e1e2}

We define two evaluation cohorts:
\begin{align}
\mathcal{P}_c^{\mathrm{E1}} &= \bigcap_{m} \mathcal{P}_{c,m}, \\
\mathcal{P}_c^{\mathrm{E2}} &= \mathcal{P}_c.
\end{align}

E1 contains only fully observed patients and serves as a clean baseline, while E2 contains all clinical patients and relies on Phase~II completion to provide missing modality representations.

To enable tensor construction, all modality latents are mapped to a shared dimensionality $D$ (set to $512$) using only deterministic zero-padding and truncation:
\begin{equation}
\bar{Z}^{(c,m)} = \mathrm{PadTrunc}\!\left(Z^{(c,m)}, D\right) \in \mathbb{R}^{N \times D},
\label{eq:align}
\end{equation}
where $\mathrm{PadTrunc}(\cdot, D)$ appends zeros when $d_m < D$ and truncates trailing dimensions when $d_m > D$. All aligned latents are standardized feature-wise prior to fusion.


\subsection{Phase 2: Hierarchical Tensor Factorization}
\label{subsec:HIT-Fuse}

The primary methodological contribution is a hierarchical tensor fusion strategy that respects biological organization by separating \emph{within-layer} integration from \emph{cross-layer} interaction modeling. Modalities are grouped according to the central dogma:
\[
D = \{\mathrm{SCNV}, \mathrm{METH}\}, \quad
R = \{\mathrm{RNA}, \mathrm{miRNA}\}, \quad
P = \{\mathrm{RPPA}, \mathrm{IMG}\}.
\]

\noindent The DNA layer (SCNV, METH) captures genomic and epigenomic regulation, the RNA layer (RNA, miRNA) captures transcriptional activity, and the functional expression layer (RPPA, IMG) captures downstream biological state. RPPA quantifies protein abundance and pathway activation, while histopathology images reflect the emergent tissue phenotype shaped by these signaling programs.
Both modalities therefore represent realized functional outcomes rather than regulatory potential \cite{Zhan2019},
justifying their integration within the same hierarchical layer.

\paragraph{Group tensor construction.}
For each group $g \in \{D,R,P\}$, we form a third-order tensor
\begin{equation}
\mathcal{X}^{(g)} = \mathrm{stack}\big(\tilde{Z}^{(c,g_1)}, \tilde{Z}^{(c,g_2)}\big) 
\in \mathbb{R}^{N \times 2 \times D},
\label{eq:group_tensor}
\end{equation}
where $N = |\mathcal{P}_c^{\mathrm{E1}}|$ or $|\mathcal{P}_c^{\mathrm{E2}}|$ and $D$ denotes the harmonized latent dimensionality.

\paragraph{CP factorization.}
Each group tensor is factorized using a rank-$R_g$ CP decomposition:
\begin{equation}
\mathcal{X}^{(g)} \approx 
\sum_{r=1}^{R_g} 
\lambda_r^{(g)} \, a_r^{(g)} \circ b_r^{(g)} \circ c_r^{(g)}.
\label{eq:cp_group}
\end{equation}

The group-level patient factors are defined as
\begin{equation}
U^{(g)} = A^{(g)} \, \mathrm{diag}(\lambda^{(g)}) 
\in \mathbb{R}^{N \times R_g}.
\label{eq:group_factors}
\end{equation}

\paragraph{Meta-tensor fusion.}
To enable cross-layer interaction modeling, the group-level representations are harmonized to a shared width 
$K_{\mathrm{meta}}$ using only zero-padding and truncation:
\begin{equation}
\bar{U}^{(g)} = \mathrm{PadTrunc}\!\left(U^{(g)}, K_{\mathrm{meta}}\right) 
\in \mathbb{R}^{N \times K_{\mathrm{meta}}}.
\end{equation}

The meta-tensor is then formed as
\begin{equation}
\mathcal{X}^{(HIT-Fuse)} = 
\mathrm{stack}\big(\bar{U}^{(D)}, \bar{U}^{(R)}, \bar{U}^{(P)}\big) 
\in \mathbb{R}^{N \times 3 \times K_{\mathrm{meta}}}.
\label{eq:meta_tensor}
\end{equation}

A second CP decomposition yields
\begin{equation}
\mathcal{X}^{(HIT-Fuse)} \approx 
\sum_{r=1}^{R_{HIT-Fuse}} 
\lambda_r^{(HIT-Fuse)} \, a_r^{(HIT-Fuse)} \circ b_r^{(HIT-Fuse)} \circ c_r^{(HIT-Fuse)}.
\label{eq:cp_meta}
\end{equation}

The final fused patient representation is
\begin{equation}
F^{(HIT-Fuse)} = A^{(HIT-Fuse)} \, \mathrm{diag}(\lambda^{(HIT-Fuse)}) 
\in \mathbb{R}^{N \times R_{HIT-Fuse}}.
\label{eq:fused}
\end{equation}

In our implementation, the shared meta width is selected deterministically from the Stage-I ranks as
\begin{equation}
K_{\mathrm{meta}} = \min\!\big(128, \max(16, \max(R_D, R_R, R_P))\big),
\end{equation}
ensuring consistent cross-layer alignment without any learned or random projection.


\subsubsection{Rank Selection via CORCONDIA and Multi-Objective Optimization}
\label{subsec:rank_selection}

For each tensor $\mathcal{X}$ (both Stage-I group tensors and the Stage-II meta-tensor), we evaluate a candidate set of ranks 
$\mathcal{R}$ and select the final rank using a two-stage criterion that balances structural validity and downstream task utility.

\paragraph{CORCONDIA screening.}
For each $r \in \mathcal{R}$, a CP decomposition is computed and the core consistency diagnostic (CORCONDIA) is evaluated, which quantifies how closely the estimated core tensor approximates an ideal superdiagonal CP core. Ranks satisfying
\begin{equation}
\mathrm{CORCONDIA}(r) \ge \tau
\end{equation}
are retained, where $\tau \in [55,60]$ depending on the tensor stage. If no rank satisfies the threshold, the top three ranks ranked by CORCONDIA are retained.

\paragraph{Multi-objective scoring.}
For each retained rank, three complementary objectives are computed:
\begin{enumerate}
\item \textbf{Predictive utility:} stratified $K$-fold cross-validated weighted F1 score using a balanced logistic regression classifier on the patient factor matrix,
\item \textbf{Survival stratification:} separation strength defined as $\mathrm{Sep}(r) = \min\!\big(50, -\log_{10}(p_r)\big)$, where $p_r$ is the log-rank test $p$-value obtained after clustering patient factors into two groups and capping survival time at 3000 days,
\item \textbf{Structural validity:} CORCONDIA itself.
\end{enumerate}

Each objective is normalized to $[0,1]$ and combined as
\begin{equation}
S(r) = w_{\mathrm{F1}} \, \tilde{F1}(r) + 
       w_{\mathrm{sep}} \, \widetilde{\mathrm{Sep}}(r) + 
       w_{\mathrm{cc}} \, \widetilde{\mathrm{CC}}(r),
\label{eq:multiobj_score}
\end{equation}
with weights $(w_{\mathrm{F1}}, w_{\mathrm{sep}}, w_{\mathrm{cc}}) = (0.55, 0.35, 0.10)$.

The selected rank is
\begin{equation}
r^{*} = \arg\max_{r \in \mathcal{R}_{\mathrm{retained}}} S(r).
\end{equation}


\subsection{Phase 3: Predictive Evaluation}
\label{subsec:downstream}

The final fused representation $F^{(HIT-Fuse)}$ and ablated baselines (M1 group-level fusion, M2 flat fusion, and modality concatenation) are evaluated on both classification and survival tasks.

\paragraph{Classification.}
We train balanced classifiers, including logistic regression, random forests, and linear support vector machines, using stratified $K$-fold cross-validation. Performance is reported using accuracy and weighted F1 score.

\paragraph{Survival analysis.}
A Cox proportional hazards model is fit to the fused representations, and concordance index (C-index) is reported. For unsupervised risk stratification, patient embeddings are clustered using agglomerative clustering, and group-wise survival differences are evaluated using the multivariate log-rank test with survival time capped at 3000 days.

\paragraph{Visualization.}
Latent spaces are visualized using t-distributed stochastic neighbor embedding (t-SNE), and Kaplan--Meier survival curves are plotted for discovered risk groups to qualitatively assess clinical separation.


\subsection{Phase 4: Explainability and Biological Validation}
\label{subsubsec:xai}

The hierarchical tensor fusion architecture enables a structured interpretability
framework that traces predictive evidence across three abstraction levels: fused
meta-factors, biological layers, and raw molecular or imaging features. We distinguish
between a \emph{global} explanation stage, which identifies cohort-wide important
tensor factors and their dominant biological drivers, and a \emph{local} explanation
stage, which produces patient-specific molecular attributions by tracing a selected
meta-factor back through the group-level tensor components and modality-specific
decoders.


\subsubsection{Global Explainability: Meta Factor and Biological Layer Importance}
\label{subsec:xai_global}

Let $F^{(HIT-Fuse)} \in \mathbb{R}^{N \times R_{HIT-Fuse}}$ denote the fused patient representation obtained from the meta level CP decomposition of the meta tensor $\mathcal{X}^{(HIT-Fuse)} \in \mathbb{R}^{N \times 3 \times K_{\mathrm{meta}}}$, where the three biological layers correspond to \emph{DNA}, \emph{RNA}, and \emph{protein--imaging (PI)} groups. In our implementation, global importance is computed for subtype classification using $F^{(HIT-Fuse)}$ as input to a downstream classifier; the same procedure applies to other tasks when an appropriate downstream model and scoring function are used.

\paragraph{Meta factor importance.}
We quantify the global importance of each meta factor $r \in \{1,\dots,R_{HIT-Fuse}\}$ using permutation importance on a downstream model trained on $F^{(HIT-Fuse)}$.
Let $\pi_r(F^{(HIT-Fuse)})$ denote the matrix obtained by permuting the $r$-th column of $F^{(HIT-Fuse)}$ across patients. We define
\begin{equation}
I_r =
\mathcal{S}\big(F^{(HIT-Fuse)}\big) -
\mathcal{S}\big(\pi_r(F^{(HIT-Fuse)})\big),
\label{eq:perm_global}
\end{equation}
where $\mathcal{S}(\cdot)$ is the downstream model score. In our code, $\mathcal{S}$ is the classifier default scoring function (accuracy for the random forest classifier); meta factors are ranked by $I_r$, and the top components are selected for driver tracing.

\paragraph{Biological layer attribution.}
From the meta level CP decomposition of $\mathcal{X}^{(HIT-Fuse)}$, the mode 2 factor matrix
$B^{(HIT-Fuse)} \in \mathbb{R}^{3 \times R_{HIT-Fuse}}$ encodes the loading of each meta factor on the three biological layers (DNA, RNA, PI). The dominant layer for meta factor $r$ is defined as
\begin{equation}
g^*(r) = \arg\max_{g \in \{\mathrm{DNA},\mathrm{RNA},\mathrm{PI}\}} \left| B^{(HIT-Fuse)}_{g,r} \right|.
\label{eq:layer_global}
\end{equation}

\paragraph{Tracing to a dominant modality within a layer.}
After selecting the dominant layer $g^*(r)$, we trace to a specific modality within that layer using the corresponding Stage I CP decomposition.
Let $B^{(g)} \in \mathbb{R}^{2 \times R_g}$ denote the Stage I mode 2 factor matrix for layer $g \in \{\mathrm{DNA},\mathrm{RNA},\mathrm{PI}\}$ with rank $R_g$.
Let $C^{(HIT-Fuse)} \in \mathbb{R}^{K_{\mathrm{meta}} \times R_{HIT-Fuse}}$ be the meta level mode 3 factor matrix. For meta factor $r$, we form a layer specific latent direction by truncating (or zero padding) the meta direction to the Stage I rank:
\begin{equation}
d_r = C^{(HIT-Fuse)}_{:,r} \in \mathbb{R}^{K_{\mathrm{meta}}}, \qquad
v_r = \mathrm{pad/trunc}(d_r, R_g) \in \mathbb{R}^{R_g}.
\end{equation}
We then score the two modalities in layer $g$ by
\begin{equation}
s_r = B^{(g)} v_r \in \mathbb{R}^{2},
\end{equation}
and choose the dominant modality as the one with maximal absolute score, i.e., $\arg\max_{j \in \{1,2\}} |(s_r)_j|$.

\paragraph{Driver decoding with integrated gradients and Jacobian projection.}
For each selected meta factor $r$, after selecting its dominant layer and modality, we obtain feature level drivers by probing the corresponding modality decoder $g_{\psi_m}$.
We compute integrated gradients (IG) in the decoder input space from a zero baseline $b=\mathbf{0}$ to the factor induced latent vector $v_r$ (after pad/trunc to the decoder input width), using the squared output energy as the attribution target:
\begin{equation}
\mathrm{IG}(v_r) = (v_r-b)\odot \mathbb{E}_{\alpha\sim[0,1]}\left[\nabla_{v}\,\left\|g_{\psi_m}\!\left(b+\alpha(v_r-b)\right)\right\|_2^2\right].
\end{equation}
To obtain attributions in the reconstructed feature space, we map the latent attribution direction through the decoder using a Jacobian vector product (JVP):
\begin{equation}
a = J_{g_{\psi_m}}(v_r)\,\mathrm{IG}(v_r),
\label{eq:ig_jvp}
\end{equation}
where $J_{g_{\psi_m}}(v_r)$ denotes the decoder Jacobian evaluated at $v_r$.
Drivers (genes, proteins, or image features) are ranked by $|a_j|$, capturing both positive and negative effects. When raw feature headers are available, we map ranked indices to the corresponding feature names; otherwise we fall back to indexed identifiers.

\paragraph{Explainability modes.}
Explainability is computed independently within each cancer cohort under two settings.
In the \emph{standard} setting, the hierarchical tensor factors are used as is.
In the \emph{balanced} setting, we apply per patient $\ell_2$ normalization to each layer level patient factor (DNA, RNA, PI) \emph{before} forming $\mathcal{X}^{(HIT-Fuse)}$ and performing the meta level CP decomposition, mitigating potential energy imbalance across biological layers for attribution purposes.
For downstream biological validation (e.g., driver inspection and enrichment analysis), we export both standard and balanced driver tables and additionally provide a merged table formed by concatenating the two exports; the merged table does not define a new attribution rule and does not perform any cross cohort merging.

\subsubsection{Local Explainability: Patient-Specific Molecular Attribution}
\label{subsec:xai_local}

Local explanations are generated for a selected patient $i$ by tracing salient meta-factors
back through the hierarchical tensor structure to raw molecular or imaging features:
\[
\text{meta-factor} \rightarrow \text{biological layer} \rightarrow \text{modality}
\rightarrow \text{latent direction} \rightarrow \text{raw features}.
\]

\paragraph{Patient-specific meta-factor selection.}
Let $F^{(HIT-Fuse)} = A_3 \odot w_3 \in \mathbb{R}^{N \times R_{HIT-Fuse}}$ denote the final patient
representation from the meta-level CP decomposition. For patient $i$, the most influential
meta-factors are selected by
\begin{equation}
\mathcal{R}_i = \mathrm{TopK}\!\left(\left|F^{(HIT-Fuse)}_{i,:}\right|\right).
\label{eq:local_topk}
\end{equation}

\paragraph{Biological layer and modality selection.}
For each selected factor $r \in \mathcal{R}_i$, the dominant biological layer is obtained
from the meta-level mode-2 loadings $B^{(HIT-Fuse)} \in \mathbb{R}^{3 \times R_{HIT-Fuse}}$:
\begin{equation}
g^*(r) = \arg\max_{g \in \{D,R,P\}} \left| B^{(HIT-Fuse)}_{g,r} \right|.
\label{eq:local_group}
\end{equation}
Let $B^{(g)} \in \mathbb{R}^{2 \times R_g}$ and $C^{(g)} \in \mathbb{R}^{D \times R_g}$ be the
group-level CP factors for layer $g^*(r)$, where $R_g$ is the Stage-I rank.

\paragraph{Local meta-direction (locality fix).}
Let $C^{(HIT-Fuse)} \in \mathbb{R}^{K_{\mathrm{meta}} \times R_{HIT-Fuse}}$ be the meta-level mode-3 factor matrix.
We form a patient-specific direction by scaling the meta direction by the patient's factor value:
\begin{equation}
a_{i,r} = F^{(HIT-Fuse)}_{i,r},
\qquad
\alpha_{i}^{(r)} = a_{i,r} \cdot C^{(HIT-Fuse)}_{1:R_{g^*(r)},\, r} \in \mathbb{R}^{R_{g^*(r)}}.
\label{eq:local_alpha}
\end{equation}
This scaling is required to avoid a global (patient-agnostic) direction and yields a
patient-specific explanation.

\paragraph{Modality attribution within the selected layer.}
The modality score within layer $g^*(r)$ is computed by projecting the group mode-2 matrix:
\begin{equation}
s_i^{(r)} = B^{(g^*(r))}\, \alpha_{i}^{(r)} \in \mathbb{R}^{2},
\qquad
m^*(r) = \arg\max_{m \in \mathrm{Mods}(g^*(r))} \left|s_{i,m}^{(r)}\right|.
\label{eq:local_mod}
\end{equation}

\paragraph{Latent direction and decoder probing.}
A factor-induced latent direction in the shared 512-d space is constructed as
\begin{equation}
z_{\mathrm{common},i}^{(r)} = C^{(g^*(r))}\, \alpha_{i}^{(r)} \in \mathbb{R}^{512}.
\label{eq:local_z}
\end{equation}
To match the true autoencoder latent dimensionality $d_{m^*(r)}$ of modality $m^*(r)$, we truncate
\begin{equation}
z_{i}^{(r,m)} = z_{\mathrm{common},i}^{(r)}[1:d_{m^*(r)}] \in \mathbb{R}^{d_{m^*(r)}}.
\label{eq:local_zslice}
\end{equation}
Finally, we decode $z_{i}^{(r,m)}$ using the
trained modality decoder $g_{\psi_m}$ and rank raw features by absolute magnitude:
\begin{equation}
\hat{x}_{i}^{(r,m)} = g_{\psi_m}\!\left(z_{i}^{(r,m)}\right) \in \mathbb{R}^{p_m},
\qquad
\text{rank by } \left|\hat{x}_{i,j}^{(r,m)}\right|.
\label{eq:local_decode}
\end{equation}
Feature indices are mapped to gene/protein identifiers or image feature channels using the original
modality feature tables.


\subsubsection{In-Silico Biological Validation and Functional Enrichment}
\label{subsubsec:bio_val}

To assess biological plausibility of model-derived molecular drivers, we perform 
independent validation using CRISPR–Cas9 gene dependency data from the Cancer 
Dependency Map (DepMap).

\paragraph{CRISPR dependency validation.}
For each cancer type $c$, we identify the set of DepMap cell lines 
$\mathcal{L}_c$ whose Oncotree lineage or primary disease annotation 
matches the corresponding TCGA cohort.
For a candidate gene $g$ present in the DepMap gene-effect matrix,
we extract gene effect scores
\[
S(g,c) = \{ s_{g,\ell} \mid \ell \in \mathcal{L}_c \}.
\]
The lineage-mean dependency is computed as
\[
\bar{s}_{g,c} = \frac{1}{|\mathcal{L}_c|} 
\sum_{\ell \in \mathcal{L}_c} s_{g,\ell}.
\]
DepMap gene-effect values are normalized such that scores near zero 
indicate non-essentiality and more negative values indicate stronger dependency.
A gene is classified as essential in cancer $c$ if
$\bar{s}_{g,c} < -0.5$.

\paragraph{Driver sets evaluated.}
Essentiality is evaluated for gene drivers obtained from the merged 
modality-specific driver exports used in the main analysis. 
For modalities that do not directly correspond to gene-level entities 
(e.g., miRNA and RPPA), validation is performed as follows.

\paragraph{RPPA validation.}
Protein targets selected from RPPA modality outputs are mapped to 
approved gene symbols and evaluated using the same DepMap dependency criterion.

\paragraph{miRNA support.}
Selected miRNAs are assessed using experimentally validated 
miRNA–target interactions from miRTarBase.
A miRNA is considered supported if at least one of its validated targets 
intersects the DepMap-evaluated gene set.

\paragraph{Functional enrichment.}
Driver gene sets are tested for over-representation in curated pathway databases (e.g.,
KEGG, Hallmark, and Gene Ontology) using a hypergeometric model with false discovery rate
correction. Enriched pathways are retained under a relaxed adjusted significance threshold
($\alpha = 0.25$) to capture weak but biologically meaningful signals in factor- and
patient-specific gene sets.

\paragraph{Integration with explainability.}
Importantly, external validation is not performed independently of the model representation but is directly anchored to the hierarchical attribution chain. 
Meta-factors are first localized to biological layers and modality branches, then decoded to ranked raw features, which are evaluated through pathway enrichment and functional screens. 
This end-to-end mapping links tensor-derived latent structure to experimentally validated biological signals, forming a closed-loop interpretability mechanism.

\section{Experiments and Results}
\subsection{Datasets and Cohort Definition}
\label{subsec:data_cohorts}

We evaluate the proposed framework on four cancer cohorts from The Cancer Genome Atlas (TCGA): glioma (GBM/LGG), lung adenocarcinoma (LUAD), breast invasive carcinoma (BRCA), and kidney cancer, comprising clear cell renal cell carcinoma (KIRC), papillary renal cell carcinoma (KIRP), and chromophobe renal cell carcinoma (KICH). For each cohort, six comics: somatic copy number variation (SCNV), DNA methylation from the HM450K platform, RNA sequencing (RNA-seq), microRNA expression (miRNA-seq), protein abundance measured by reverse-phase protein arrays (RPPA), and digital pathology features extracted from hematoxylin and eosin (H\&E) stained whole-slide images (WSIs). The omics features are obtained from LinkedOmics \cite{vasaikar2018linkedomics}, while the WSI-derived pathomics features are based on self-supervised vision transformer representations following \cite{hussein2026vit}. Clinical variables include overall survival time, event status, tumor histological subtype or risk group.
Modality-specific availability defines the two evaluation configurations used in this study. The real-only intersection cohort (E1) contains patients with complete availability across all six modalities and clinical annotations. In contrast, the completed all-patient cohort (E2) includes patients with clinical data and at least one observed modality, and missing modality embeddings are estimated via missing-modality latent reconstruction. This design allows us to compare performance under a strict complete-case setting and a more realistic setting that reflects incomplete modality availability in large clinical cohorts.

All modality-specific latent representations are fixed before tensor fusion. SCNV and DNA methylation are encoded into 512-dimensional latent vectors, RNA-seq, miRNA-seq, and image features are encoded into 256-dimensional latent vectors, and RPPA is encoded into a 64-dimensional latent vector. These latent representations form the input to the biologically structured hierarchical tensor factorization stage described in the following sections.

\begin{table}[t]
\centering
\footnotesize
\setlength{\tabcolsep}{3pt}

\begin{minipage}[t]{0.49\linewidth}
\centering
\caption{Cohort sizes and expansion from E1 to E2.}
\label{E1E2cohort}
\begin{tabular}{lccc}
\toprule
Cancer & E1 & E2 & Expansion ($\times$) \\
\midrule
Glioma & 397 & 1110 & 2.80 \\
Breast & 501 & 1097 & 2.18 \\
Lung   & 283 & 522  & 1.84 \\
Kidney & 408 & 941  & 2.31 \\
\bottomrule
\end{tabular}
\end{minipage}
\hfill
\begin{minipage}[t]{0.49\linewidth}
\centering
\caption{Selected hierarchical tensor ranks across cohorts.}
\label{tensor-rank}
\begin{tabular}{lcccc}
\toprule
Cancer & Stage-D & Stage-R & Stage-P & Meta \\
\midrule
BRCA   & 4 & 8 & 8 & 8 \\
Glioma & 8 & 8 & 8 & 8 \\
Kidney & 8 & 8 & 8 & 4 \\
LUAD   & 8 & 8 & 2 & 4 \\
\bottomrule
\end{tabular}
\end{minipage}

\end{table}

\noindent Table~\ref{E1E2cohort} summarizes cohort sizes under both configurations. 
Across the studied cancers. E2 substantially increases the number of analyzable patients relative to E1.

\subsection{Structural Stability of the Hierarchical Tensor Decomposition}
\label{sec:rank_stability}

We evaluate the structural behavior of the hierarchical tensor decomposition across four cancer cohorts. Rank selection follows the CORCONDIA-guided multi-objective strategy described in Section~3.6; here we report the final selected ranks. Table~\ref{tensor-rank} presents the chosen Stage-I ranks for the DNA, RNA, and Protein–Imaging groups, together with the Stage-II meta rank. Stage-I ranks vary across biological subsystems and cancer types, indicating heterogeneous modality complexity. 
In contrast, the meta-level rank remains moderate across cohorts ($R_{\mathrm{meta}} \in \{4,8\}$). 
Kidney and LUAD converge to a meta rank of $4$, whereas BRCA and Glioma select $8$. Importantly, no cohort requires high-rank expansion in either Stage-I or Stage-II decomposition. The selected configurations remain bounded and consistent across cancers, indicating stable hierarchical factorization under the proposed tuning procedure.

\begin{figure}
    \centering
    \includegraphics[width=1\linewidth]{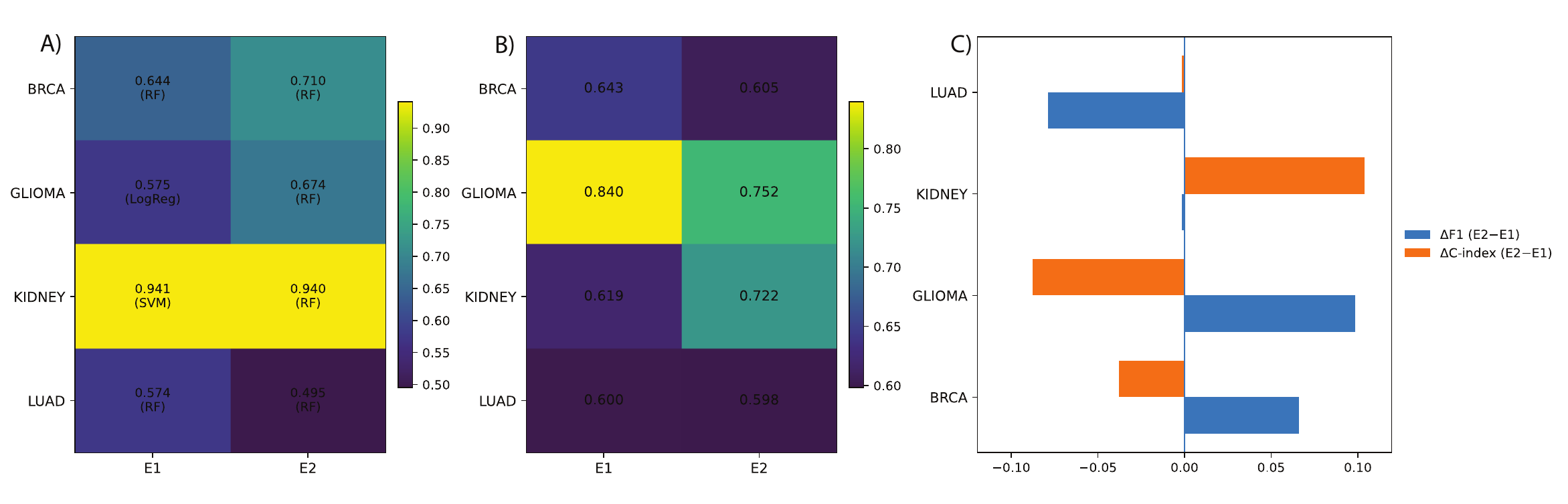}
    \caption{
Predictive and prognostic validation of the HIT-Fuse hierarchical tensor framework.
(A) Best subtype classification performance (Weighted F1) per cancer under E1 and E2, with the dominant classifier indicated in parentheses.
(B) Survival discrimination measured by C-index from the Cox survival head.
(C) Effect of cohort completion (E2$-$E1) on subtype and survival metrics, illustrating cancer-specific stability patterns.
}
\label{fig:HIT-Fuse_validation}
\end{figure}

\subsection{Predictive and Prognostic Validation of our Proposed Hierarchical Tensor Framework}
\label{sec:HIT-Fuse_validation}
\begin{figure*}[t]
\centering

\includegraphics[width=\textwidth]{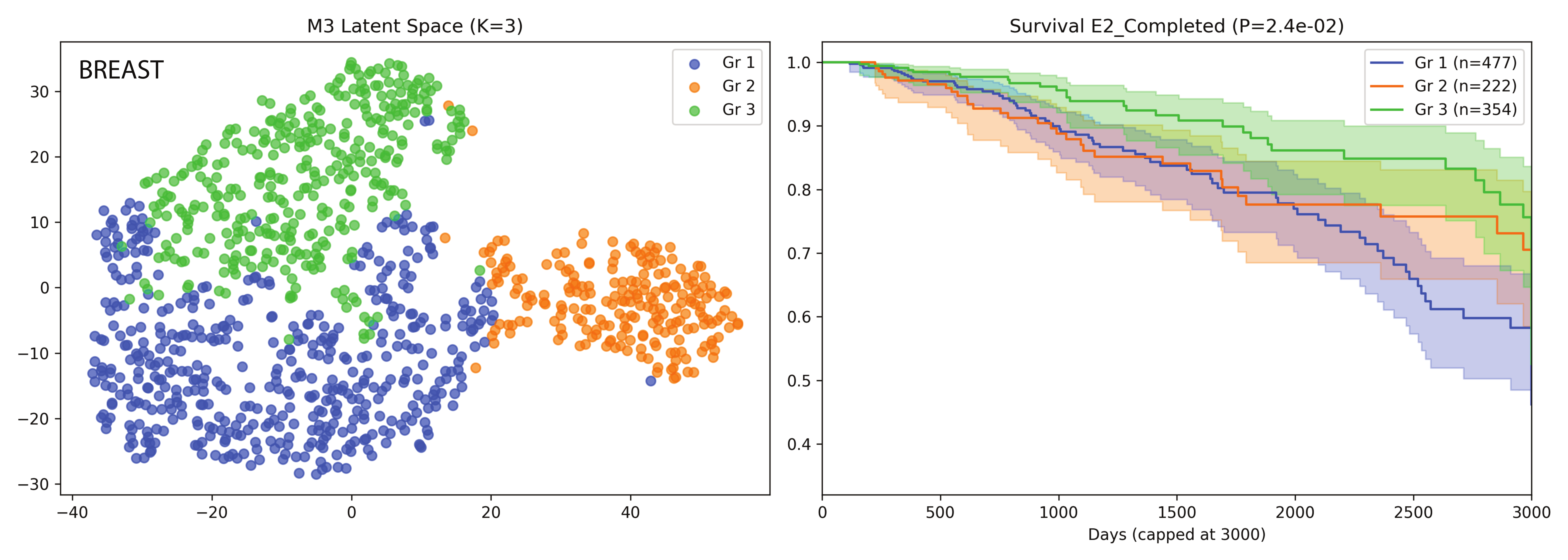}

\vspace{0.6em}

\includegraphics[width=\textwidth]{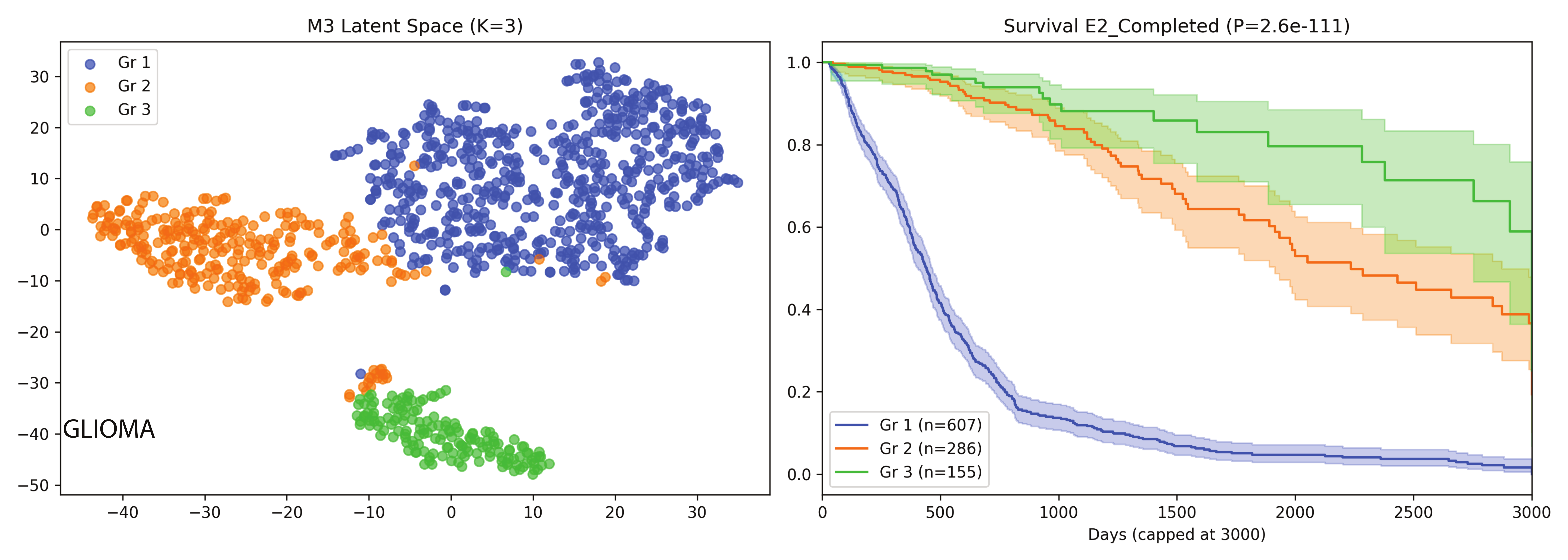}

\vspace{0.6em}

\includegraphics[width=\textwidth]{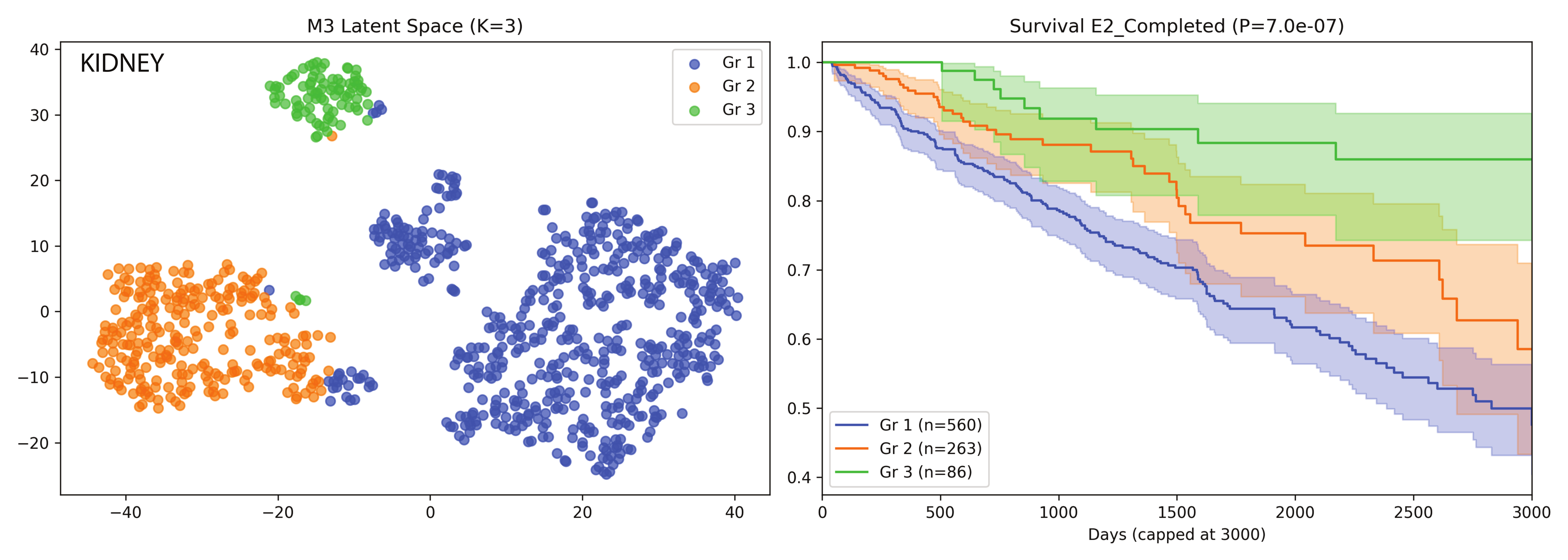}

\vspace{0.6em}

\includegraphics[width=\textwidth]{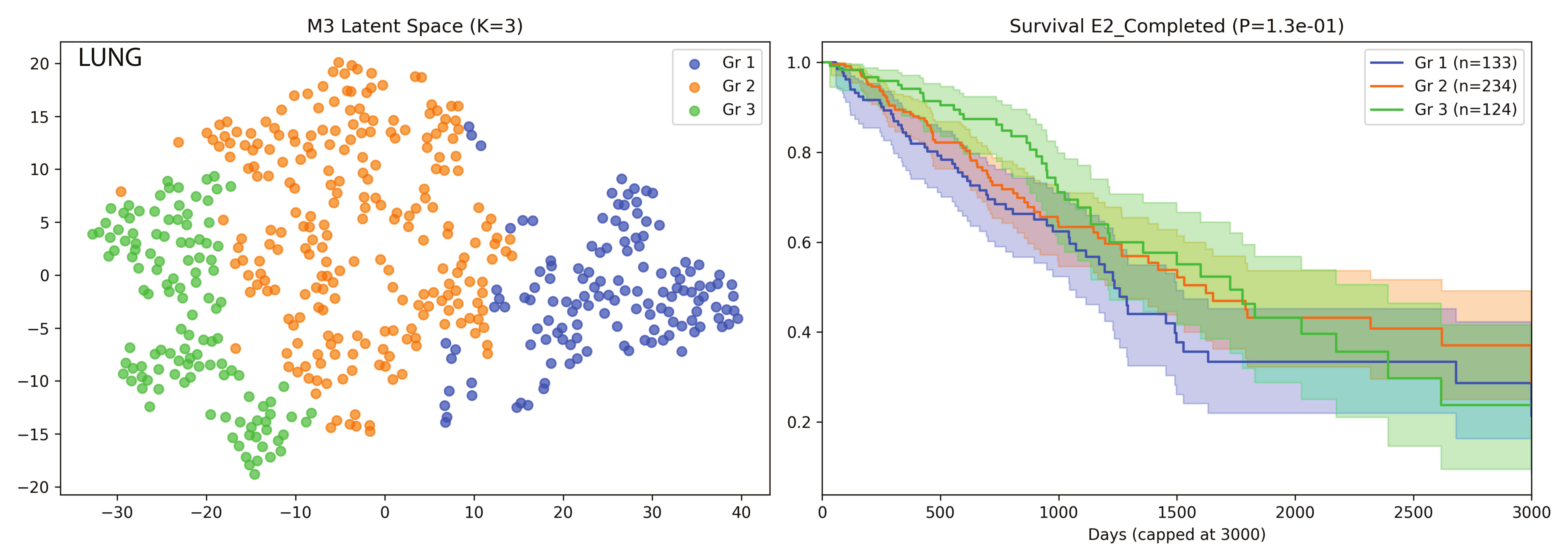}

\caption{\textbf{Latent-space clustering and survival stratification under HIT-Fuse (E2, $K{=}3$).}
For each cohort, patients are clustered in the learned latent space (left within each panel) and evaluated via Kaplan--Meier survival separation (right). Survival is truncated at 3000 days for comparability across cohorts.}
\label{fig:HIT-Fuse_k3_grid}
\end{figure*}

The predictive and clinical utility of the proposed hierarchical tensor framework (HIT-Fuse) is evaluated across four TCGA cohorts in two settings: E1 (intersection) and E2 (completed). Performance is assessed along two complementary axes: subtype classification and survival prognosis. Subtype prediction is evaluated using logistic regression (LR), random forest (RF), and linear SVM, with results reported as Accuracy and weighted F1 (F1(w)). Survival discrimination is quantified using the concordance index (C-index) from a Cox proportional hazards model. Clinical stratification is further examined by clustering the HIT-Fuse latent space and evaluating Kaplan–Meier separation with log-rank testing, with survival truncated at 3000 days for comparability. Figure~\ref{fig:HIT-Fuse_validation} summarizes the principal trends, In Panel (A), the strongest subtype performance under E2 is achieved by RF for BRCA (F1(w)=0.710), RF for GLIOMA (0.674), RF for KIDNEY (0.940), and RF/SVM jointly for LUAD (0.495), indicating that tree-based models consistently dominate in the enriched cohort. Panel (B) shows that survival discrimination remains stable under E2, with C-index values of 0.605 (BRCA), 0.752 (GLIOMA), 0.722 (KIDNEY), and 0.598 (LUAD). Panel (C) illustrates that cohort completion (E2–E1) produces cancer-specific shifts rather than uniform improvements, with performance gains observed in several subtype metrics and stable survival discrimination across cohorts. Table~\ref{tab:HIT-Fuse_full_classifier_breakdown} provides the complete classifier-level breakdown for Accuracy, F1(w), and C-index across cohorts and settings. Collectively, the results indicate that our framework maintains competitive subtype performance and stable survival discrimination while benefiting from increased sample size, supporting the robustness of the hierarchical tensor design.
Figure~\ref{fig:HIT-Fuse_k3_grid} evaluates survival stratification in the HIT-Fuse latent space under the completed setting (E2) with $K{=}3$ clusters. For each cohort, the left panel shows a two-dimensional t-SNE projection of the learned latent representations, while the right panel presents Kaplan–Meier survival curves derived from cluster assignments. Significant survival separation is observed for Breast ($p = 2.4\times10^{-2}$), Glioma ($p = 2.6\times10^{-111}$), and Kidney ($p = 7.0\times10^{-7}$), indicating that latent clusters correspond to distinct risk groups. In LUAD, separation is weaker ($p \approx 1.3\times10^{-1}$), despite a cohort size exceeding 500 patients under E2. Overall, the t-SNE visualization and survival analysis show that the HIT-Fuse latent representation supports statistically significant stratification in most cohorts without evidence of instability following cohort expansion.

\begin{figure}[t]
\centering
\vspace{-2mm}
\includegraphics[width=0.92\linewidth]{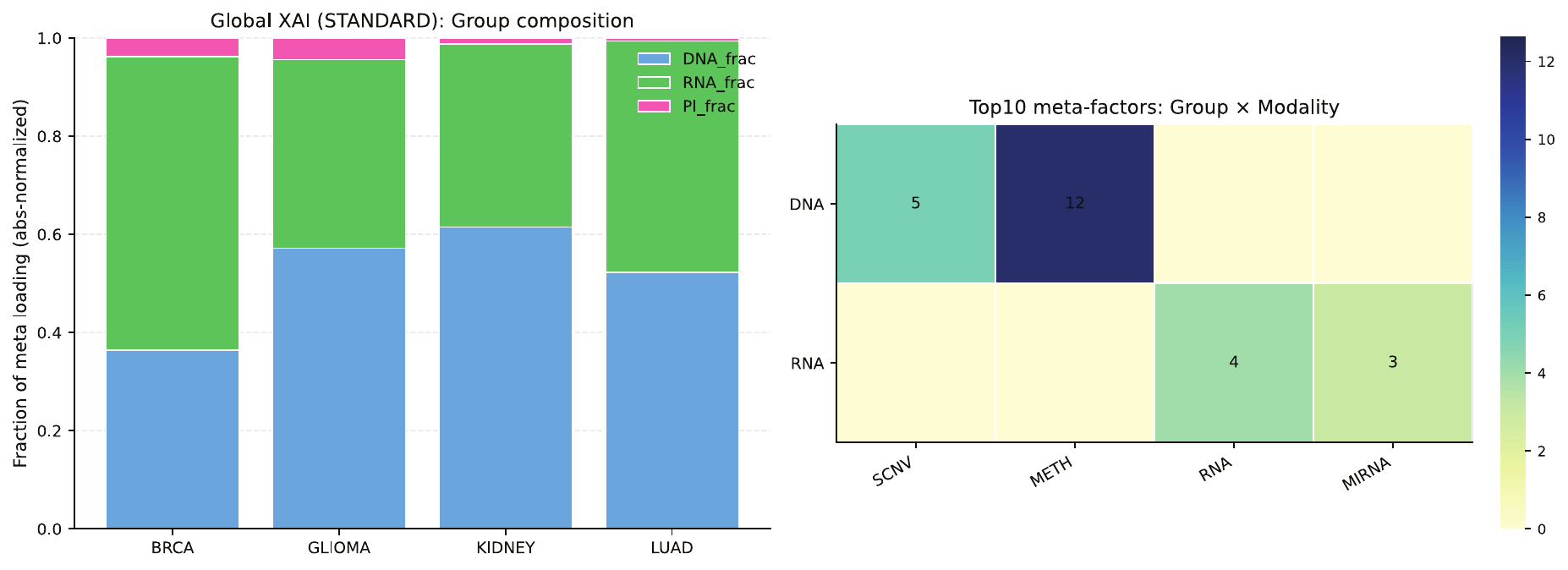}
\vspace{-2mm}
\caption{\textbf{Global explainability under the standard setting.}
Left: Fractional contribution of biological layers (DNA, RNA, protein--imaging) from absolute meta-level loadings $B^{(HIT-Fuse)}$ aggregated over meta-factors and normalized per cohort.
Right: Frequency of top-ranked meta-factors across group--modality combinations, pooled across cohorts (top-10 per cohort ranked by permutation importance).}
\label{fig:xai123}
\vspace{-2mm}
\end{figure}


\begin{table*}[t]
\centering
\scriptsize
\renewcommand{\arraystretch}{1.1}
\setlength{\tabcolsep}{3pt}

\caption{
Detailed HIT-Fuse hierarchical tensor framework performance across subtype classifiers and survival discrimination.
The best F1(w) per cohort is highlighted in bold. C-index corresponds to the survival head.
}
\label{tab:HIT-Fuse_full_classifier_breakdown}

\resizebox{\textwidth}{!}{%
\begin{tabular}{lccccccc ccccccc}
\toprule
& \multicolumn{7}{c}{\textbf{E1}}
& \multicolumn{7}{c}{\textbf{E2}} \\
\cmidrule(lr){2-8} \cmidrule(lr){9-15}
\textbf{Cancer}
& \multicolumn{3}{c}{Accuracy}
& \multicolumn{3}{c}{F1(w)}
& C
& \multicolumn{3}{c}{Accuracy}
& \multicolumn{3}{c}{F1(w)}
& C \\
\cmidrule(lr){2-4} \cmidrule(lr){5-7}
\cmidrule(lr){9-11} \cmidrule(lr){12-14}
& LR & RF & SVM & LR & RF & SVM &
& LR & RF & SVM & LR & RF & SVM & \\
\midrule

BRCA
& 0.417 & 0.701 & 0.677
& 0.474 & \textbf{0.644} & 0.596
& 0.643
& 0.421 & 0.764 & 0.767
& 0.502 & \textbf{0.710} & 0.708
& 0.605 \\

GLIOMA
& 0.590 & 0.584 & 0.597
& \textbf{0.575} & 0.567 & 0.519
& 0.840
& 0.508 & 0.716 & 0.699
& 0.561 & \textbf{0.674} & 0.652
& 0.752 \\

KIDNEY
& 0.941 & 0.934 & 0.941
& \textbf{0.941} & 0.934 & \textbf{0.941}
& 0.619
& 0.914 & 0.939 & 0.922
& 0.918 & \textbf{0.940} & 0.924
& 0.722 \\

LUAD
& 0.319 & 0.646 & 0.635
& 0.387 & \textbf{0.574} & 0.494
& 0.600
& 0.183 & 0.595 & 0.636
& 0.236 & \textbf{0.495} & \textbf{0.495}
& 0.598 \\

\bottomrule
\end{tabular}%
}
\end{table*}

\subsection{Global and Local Explainability}

Figure~\ref{fig:xai123} summarizes cohort-level global explainability under the standard attribution setting. 
Meta-factors are ranked according to permutation importance computed on the final meta representation $F^{(HIT-Fuse)}$, where importance reflects the mean performance drop of a random forest classifier under feature permutation. 
The left panel reports the relative contribution of each biological group by aggregating the absolute meta-level group loadings $B^{(HIT-Fuse)}$ across meta-factors. 
The right panel pools the top-10 ranked meta-factors from all cohorts and displays the frequency of their (group, modality) assignments, where group dominance is determined by $\arg\max_g |B^{(HIT-Fuse)}_{g,r}|$ and modality selection follows projection onto the corresponding stage-1 modality loadings. Table~\ref{tab:representative_meta} reports one representative meta-factor per cohort, defined as the highest-ranked meta-factor under the same permutation-importance criterion. 
For each cohort, we list its dominant biological group, selected modality, and the top decoded drivers (top five shown for brevity.

\begin{table}[t]
\centering
\setlength{\tabcolsep}{5pt}
\renewcommand{\arraystretch}{1.15}
\small
\begin{tabular}{l c c c l}
\hline
Cancer & Latent Component & Group & Modality & Top Drivers (Top 5) \\
\hline
BRCA 
& 1 & RNA & MIRNA 
& \makecell[l]{hsa-mir-1322, hsa-mir-518a-2, hsa-mir-518b,\\
hsa-mir-670, hsa-mir-519a-2} \\

GLIOMA 
& 4 & DNA & METH 
& \makecell[l]{ZNF544, XRCC5, STK16, GATA5,\\
CPLX2} \\

KIDNEY 
& 2 & DNA & METH 
& \makecell[l]{RAD51AP2, ISYNA1, VANGL2,\\
SLC7A6, ARHGDIG} \\

LUAD 
& 1 & DNA & METH 
& \makecell[l]{ENO3, STRA8, WRB,\\
ZNF286A, ANXA5} \\

\hline
\end{tabular}
\caption{Representative highest-importance meta-factor per cancer cohort under the standard setting. Only the top five decoded drivers are shown for brevity.}
\label{tab:representative_meta}
\end{table}

\paragraph{Local explainability: representative patient case studies.}
The proposed local traceback mechanism can be applied to any individual patient within each cancer cohort. 
To illustrate its behavior, we present representative case studies from the kidney and glioma cohorts. 
For a given patient $p$, we select the top meta-factors by magnitude in $F^{(HIT-Fuse)}_{p,:}$, trace each factor to its dominant biological layer (DNA, RNA, or PI) via $B^{(HIT-Fuse)}$, scale the meta direction by the patient loading $\alpha_{p,f}=F^{(HIT-Fuse)}_{p,f}$ to ensure locality, and decode the resulting modality-specific latent direction through the corresponding autoencoder decoder to obtain ranked driver features.

\textbf{Kidney (TCGA.GL.A9DE).}
The model predicted papillary renal cell carcinoma with low-risk status. 
The highest-magnitude meta-factor ($f=0$, $\alpha_{p,f}=11.21$) mapped to the DNA layer with dominant modality methylation, producing drivers including \emph{LSM4}, \emph{TRIM60}, \emph{VHL}, and \emph{POM121L4P}. 
The second-ranked factor ($f=1$, $\alpha_{p,f}=7.85$) traced to the DNA layer via SCNV, yielding drivers such as \emph{ATP2B4}, \emph{KISS1}, and \emph{SNORD112}. 
A third factor ($f=2$, $\alpha_{p,f}=-1.40$) again mapped to methylation with drivers including \emph{PACS1}, \emph{DLX2}, and \emph{GALNS}. 
These results show that distinct meta-factors within the same patient correspond to different molecular modalities, reflecting multiple biological axes captured by the fused representation.

\textbf{Glioma (TCGA.02.0004).}
For a patient predicted as untreated primary (de novo) GBM, the top meta-factors were strongly associated with the DNA layer but separated into modality-specific contributions. 
Factor $f=3$ ($\alpha_{p,f}=19.20$) mapped to methylation with drivers including \emph{MEF2B}, \emph{DSN1}, \emph{SHANK3}, and \emph{NP1X1}. 
Factor $f=5$ ($\alpha_{p,f}=-12.87$) also mapped to methylation with drivers such as \emph{MLF1}, \emph{SHISA3}, and \emph{DOCK6}. 
In contrast, factor $f=6$ ($\alpha_{p,f}=-8.36$) mapped to SCNV, producing drivers including \emph{GFRA1}, \emph{CGN}, and \emph{PPP5K1A}. 
Thus, even when the dominant biological layer remains the same, the local explanation differentiates between copy-number and methylation axes in a patient-specific manner.
For patients whose dominant modality was imaging (PI layer, IMG branch), the decoded top features correspond to latent dimensions derived from whole-slide image patch embeddings and features can be mapped back to their originating patches, enabling spatial interpretation of the explanation.  These examples demonstrate that the proposed traceback mechanism yields patient-specific, modality-resolved driver patterns rather than a single global driver direction.

\subsection{Biological Interpretation and External Validation}

\begin{figure*}[t]
\centering

\includegraphics[width=\textwidth]{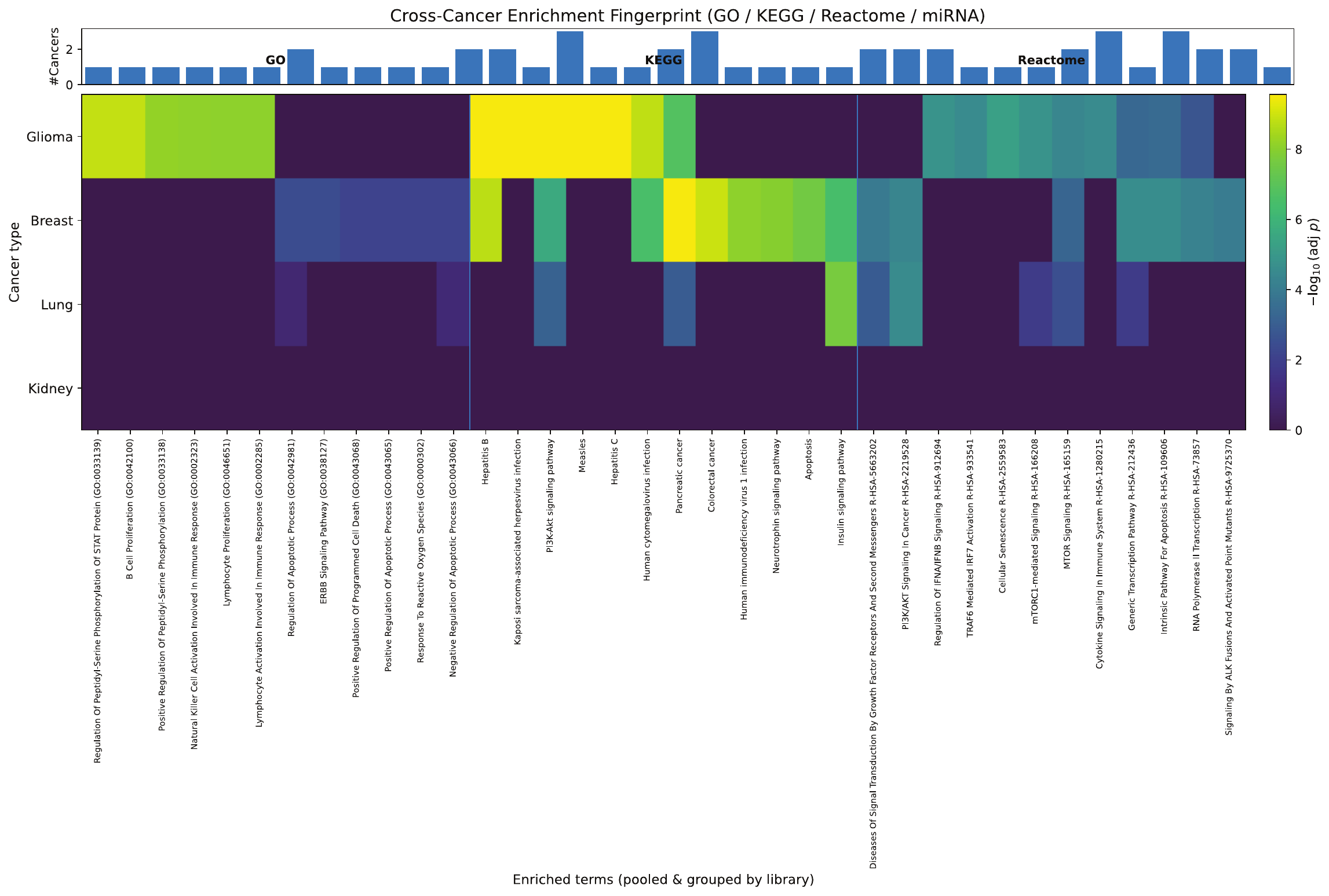}

\vspace{1em}

\includegraphics[width=\textwidth]{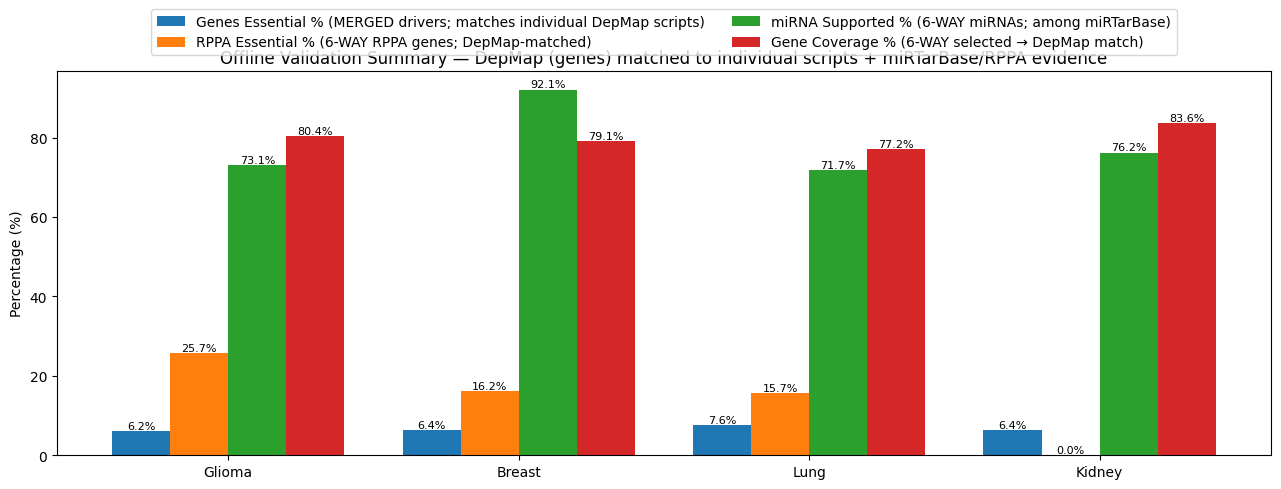}

\caption{
\textbf{Biological interpretation and external validation of HIT-Fuse-derived drivers.}
\textbf{Top A:} Cross-cancer enrichment fingerprint across GO Biological Process, KEGG, Reactome, miRTarBase, and miEAA libraries. 
Heatmap intensity reflects $-\log_{10}(\mathrm{adjusted}\ p)$, and the upper bar indicates the number of cancers in which each term is significantly enriched.
\textbf{Bottom B:} External functional validation using DepMap CRISPR essentiality screens and regulatory evidence. 
Bars indicate overall DepMap gene coverage, essential gene proportion, RPPA-supported genes, and miRTarBase-supported miRNAs.
}
\label{fig:biological_validation}
\end{figure*}
\subsubsection{Cross-Cancer Pathway Enrichment Analysis}
\label{subsec:enrichment}
To interpret the biological mechanisms captured by the HIT-Fuse framework, we performed functional enrichment analysis on the union of important genes derived from meta-factor drivers in each cohort. Enrichment was evaluated using Gene Ontology (GO), KEGG, Reactome, and experimentally supported miRNA target libraries. Figure~\ref{fig:biological_validation}A presents a cross-cancer enrichment fingerprint summarizing significantly enriched pathways across glioma, breast, lung, and kidney cohorts. Enrichment intensity is reported as $-\log_{10}(\mathrm{adjusted}\ p)$. In glioma, enrichment was strongly concentrated in immune-related processes, apoptotic regulation, and PI3K–AKT signaling, consistent with known glioblastoma immune and survival dependencies. Breast cancer exhibited enrichment in cell cycle regulation, apoptosis, and receptor tyrosine kinase pathways. Lung cancer showed moderate enrichment in immune signaling and apoptotic pathways. In contrast, kidney displayed comparatively fewer significantly enriched pathways in the pooled export, reflecting cohort-specific driver sparsity. Across cancers, recurrent enrichment in apoptosis, immune signaling, PI3K–AKT pathway, and cytokine signaling suggests that the HIT-Fuse-derived drivers consistently capture core oncogenic axes while preserving cancer-specific variation.

\subsubsection{External Functional Validation Using DepMap CRISPR Screens}
\label{subsec:depmap_validation}
To further assess functional relevance, we evaluated selected drivers against 
DepMap CRISPR–Cas9 gene dependency screens. As summarized in Fig.~\ref{fig:biological_validation}, 
HIT-Fuse-derived drivers demonstrate broad coverage within DepMap gene-effect matrices, 
with 77.2–83.6\% of selected genes matched across cohorts. Among DepMap-matched genes, 6–8\% satisfied the essentiality criterion 
($\bar{s}_{g,c} < -0.5$) in lineage-matched cell lines. Representative 
essential drivers included RPL14 and PRPF31 (kidney), SNRPD1 and RPL6 (breast), 
PLK1 and DHODH (lung), and ANAPC11 and PSMB4 (glioma). In addition, 71.7–92.1\% of cohort-specific miRNAs were supported by 
experimentally validated miRTarBase targets, while RPPA-selected genes 
exhibited variable essentiality rates (0–25.7\%) across cohorts. 
Together, these results indicate that HIT-Fuse-identified drivers are not only 
statistically salient within the tensor fusion space but also enriched 
for experimentally supported functional dependencies.

\subsection{Fusion Architecture Ablation Study}
\label{subsec:ablation}

To isolate the effect of \emph{fusion design}, we ablate the fusion architecture while keeping all other components fixed: (i) the same modality-specific latent representations, (ii) the same CORCONDIA-guided rank tuning procedure, and (iii) the same downstream subtype classifiers. Thus, any differences reflect the inductive bias introduced by the fusion mechanism rather than changes in feature extraction or tuning.

\paragraph{Ablation settings.}
We compare four fusion strategies:
\begin{itemize}
    \item \textbf{Concat\_All (feature concatenation).} All modality latents are concatenated into a single vector per patient and fed directly to the classifier. This baseline imposes no explicit cross-modal structure.
    \item \textbf{M1\_GroupCP (group-wise CP).} Modalities are first grouped according to a coarse biological flow: DNA (SCNV+METH), RNA (RNA-seq+miRNA), and protein--imaging (RPPA+IMG). A CP decomposition is applied \emph{separately} to each group tensor to obtain group-specific patient factors, which are then concatenated for prediction. This enforces within-group interaction modeling but does not explicitly model interactions \emph{between} groups.
    \item \textbf{M2\_FlatCP (single-stage CP).} All modalities are stacked into a single tensor and decomposed with CP, producing one set of patient factors that captures interactions across all modalities in a single stage, without biological grouping.
    \item \textbf{HIT-Fuse (proposed hierarchical fusion).} HIT-Fuse extends M1 by introducing a second-stage meta-factor decomposition over the three group representations (DNA/RNA/protein--imaging), explicitly capturing cross-group interactions while preserving the biological grouping.
\end{itemize}

\paragraph{Quantitative results.}
Table~\ref{tab:ablation_full} summarizes the \emph{maximum weighted F1} achieved by each fusion strategy in each cancer--cohort setting (E1/E2), where the reported value for each fusion corresponds to its best-performing classifier among the evaluated subtype heads.

\begin{table}[t]
\centering
\caption{Maximum weighted F1 scores achieved under each fusion strategy across cancers and cohorts. For each fusion method, the best-performing subtype classifier (among evaluated models) is reported. Bold indicates the highest F1 within each cancer--cohort setting.}
\label{tab:ablation_full}
\begin{tabular}{llcccc}
\toprule
Cancer & Cohort & Concat\_All & M1\_GroupCP & M2\_FlatCP & HIT-Fuse \\
\midrule
Breast & E1 & \textbf{0.736} & 0.666 & 0.617 & 0.644 \\
Breast & E2 & \textbf{0.775} & 0.716 & 0.657 & 0.710 \\
Glioma & E1 & 0.551 & 0.574 & \textbf{0.583} & 0.575 \\
Glioma & E2 & \textbf{0.726} & 0.698 & 0.688 & 0.674 \\
Kidney & E1 & \textbf{0.973} & 0.958 & 0.921 & 0.941 \\
Kidney & E2 & \textbf{0.960} & 0.941 & 0.924 & 0.940 \\
LUAD   & E1 & 0.561 & 0.553 & 0.517 & \textbf{0.574} \\
LUAD   & E2 & \textbf{0.536} & 0.500 & 0.495 & 0.495 \\
\bottomrule
\end{tabular}
\end{table}

Across the eight cancer--cohort combinations, \textbf{Concat\_All} attains the highest weighted F1 in six settings, indicating that strong subtype-discriminative signal is already present in the aggregated latent features and can be exploited by standard classifiers without imposing additional structure. Nonetheless, tensor fusion is advantageous in specific cohorts: \textbf{M2\_FlatCP} achieves the top result in Glioma (E1), and \textbf{HIT-Fuse} achieves the top result in LUAD (E1), suggesting that structured interaction modeling can be beneficial depending on cohort characteristics and tumor-specific cross-modal relationships. Importantly, no single fusion strategy dominates universally, highlighting that fusion inductive biases interact with cohort composition (E1 vs.\ E2) and cancer biology.

\paragraph{Interpretability implication.}
While HIT-Fuse is not the top performer in weighted F1 in most settings, it remains competitive and is uniquely designed to support \emph{mechanistic interpretation} by construction. In particular, HIT-Fuse enables a \emph{more direct and biologically aligned} patient-level traceback of drivers through staged factorization (meta-factor $\rightarrow$ biological layer $\rightarrow$ modality branch $\rightarrow$ decoded drivers), whereas flat concatenation and single-stage tensor fusion do not \emph{natively} impose this hierarchy and therefore require heavier post-hoc assumptions to obtain comparable mechanistic narratives. In this sense, the ablation indicates that predictive optimization (often favoring concatenation) and biologically structured fusion (supporting coherent cross-modal interpretation) are complementary objectives, and HIT-Fuse targets the latter while maintaining competitive predictive performance.

\section{Discussion}
\label{sec:results_discussion}

\paragraph{Cohort Coverage and Tensor Stability.} Across four TCGA cohorts and two cohort constructions (E1 intersection vs.\ E2 completed), the experiments highlight a consistent theme: \emph{predictive optimality and biological structuring are not identical objectives}. E2 systematically increases cohort coverage (Table~\ref{E1E2cohort}), enabling evaluation on substantially larger and more clinically representative patient sets. This expansion is aligned with the practical reality of multimodal clinical cohorts, where complete modality availability is uncommon and missingness is often structured rather than random~\cite{tcga2013pancancer,vasaikar2018linkedomics}. Importantly, the hierarchical decomposition remains well behaved under expansion: tuned ranks remain moderate and bounded across cancers (Table~\ref{tensor-rank}), suggesting that the CORCONDIA-guided, multi-objective procedure can avoid unstable over-parameterization while adapting to cohort-specific complexity~\cite{bro2003corcondia,kolda2009tensor}.

\paragraph{Predictive and prognostic behavior under cohort expansion.}
Under HIT-Fuse, subtype performance varies non-uniformly from E1 to E2 (Fig.~\ref{fig:HIT-Fuse_validation}A,C), while survival discrimination remains comparatively stable (Fig.~\ref{fig:HIT-Fuse_validation}B). This pattern is plausible in multimodal oncology settings: increasing sample size and heterogeneity can alter subtype separability (especially for imbalanced or clinically heterogeneous labels), whereas survival modeling often benefits from larger cohorts and may be less sensitive to marginal class-boundary shifts when trained with regularized Cox objectives~\cite{cox1972regression,harrell1982cindex}. The strong KM separations observed in BRCA, Glioma, and Kidney under E2 (Fig.~\ref{fig:HIT-Fuse_k3_grid}) further indicate that the learned fused representation contains clinically meaningful axes that translate into risk stratification, consistent with prior multimodal survival literature combining pathology and omics signals~\cite{ValeSilva2021MultiSurv}.

\paragraph{Why concatenation often dominates the ablation table.}
The fusion ablation (Table~\ref{tab:ablation_full}) shows that \textbf{Concat\_All} yields the highest weighted F1 in most cancer--cohort settings. This is not necessarily surprising in your pipeline because each modality is already represented by a learned latent embedding (fixed-dimensional, standardized, and information-dense). When strong predictive signal is already present in the latent space, a linear or tree-based classifier on concatenated features can be highly effective, and may outperform structured tensor objectives that impose additional constraints on interactions. In other words, concatenation can be viewed as an \emph{upper baseline} for pure predictive optimization when interpretability constraints are not enforced.

\paragraph{When tensor structure helps (and what it buys you even when it does not win F1).}
Despite concatenation’s dominance, tensor-based strategies provide two advantages that matter for the scientific aims of this work. First, they explicitly model \emph{multiway interactions} rather than treating modalities as an unordered feature bag, which can be beneficial when prognostic signal is expressed through cross-modality coupling (e.g., coordinated DNA--RNA--protein patterns)~\cite{kolda2009tensor}. This is consistent with settings where M2 or HIT-Fuse is competitive or best (e.g., Glioma E1 and LUAD E1 in Table~\ref{tab:ablation_full}). Second, and more importantly for the paper’s contributions, the hierarchical design induces a \emph{structured factorization aligned with biological flow} (DNA $\rightarrow$ RNA $\rightarrow$ protein/imaging), which supports coherent interpretation. In HIT-Fuse, each patient’s meta-factor activations can be traced to a dominant biological layer via meta loadings and then down to a specific modality branch (and ultimately decoded to ranked drivers), yielding a reproducible chain of attribution. Flat concatenation does not provide this decomposition natively, and single-stage tensor fusion (M2) does not explicitly separate within-layer vs.\ cross-layer interactions, making such traceback less direct and less biologically anchored.

\paragraph{Biological validity signals from enrichment and functional screens.}
The enrichment fingerprint (Fig.~\ref{fig:biological_validation}A) indicates that driver exports capture both shared oncogenic programs and cancer-specific signatures, and uses established functional libraries to contextualize the decoded drivers. External validation using DepMap CRISPR dependency profiles (Fig.~\ref{fig:biological_validation}B) further supports functional relevance: mapping drivers to CRISPR target genes and observing substantial essentiality rates in cancer-matched cell lines is consistent with the Dependency Map paradigm for identifying cancer vulnerabilities~. The miRNA support via experimentally validated targets (miRTarBase) provides complementary regulatory plausibility for miRNA-linked drivers. While these validations do not prove causality, they strengthen the claim that HIT-Fuse’s decoded drivers are not arbitrary latent artifacts but correspond to experimentally interrogated functional units.

\paragraph{Limitations and practical takeaway.}
The results suggest that \emph{no single fusion strategy universally dominates} across cancers, cohorts, and endpoints. If the sole objective were maximizing subtype F1, Concat\_All is often the strongest choice in this experimental design. However, the central goal of HIT-Fuse is to balance competitive performance with \emph{biologically structured, mechanistically interpretable fusion}. Thus, the main takeaway is not that HIT-Fuse is always the top classifier, but that it provides a principled representation that (i) remains stable under cohort completion, (ii) supports clinically meaningful stratification, and (iii) enables patient-level, layer-resolved driver tracing with downstream biological validation (enrichment and DepMap), properties that are harder to obtain in a single step from purely predictive baselines.

\section{Conclusion}

We introduced HIT-Fuse, a hierarchical tensor-based multimodal fusion framework designed to integrate genomics, transcriptomics, proteomics, and imaging representations while preserving biologically motivated structure. Unlike flat concatenation or single-stage tensor approaches, HIT-Fuse explicitly separates within-layer interactions (DNA, RNA, protein–imaging) from cross-layer meta interactions, enabling structured factorization aligned with biological flow.

Across four TCGA cancer cohorts and two cohort constructions (intersection E1 and completed E2), HIT-Fuse demonstrated competitive subtype classification performance and stable survival discrimination under cohort expansion. Although direct feature concatenation frequently achieved the highest weighted F1, HIT-Fuse remained consistently competitive while uniquely supporting layer-resolved, patient-specific driver traceback. Cohort expansion through latent completion substantially increased sample coverage without destabilizing hierarchical rank selection or degrading prognostic stratification.

Biological interpretability analysis further showed that HIT-Fuse-derived drivers recapitulate known oncogenic processes across cancers and are enriched for experimentally validated functional dependencies, including DepMap CRISPR gene essentiality profiles and curated miRNA target interactions. These findings indicate that the hierarchical representation captures biologically meaningful cross-modal structure rather than arbitrary latent correlations.

Overall, this study highlights an important distinction between predictive optimization and biologically structured integration. While concatenation may maximize classification accuracy in certain settings, structured multimodal factorization provides a principled foundation for mechanistic interpretation and cross-modal reasoning. The proposed framework therefore offers a scalable and interpretable strategy for multimodal cancer modeling in heterogeneous clinical cohorts.

\clearpage
\section*{Declarations}

\subsection*{Ethics approval and consent to participate}
Not applicable.

\subsection*{Consent for publication}
Not applicable.

\subsection*{Availability of data and materials}
The datasets used in this study are publicly available from The Cancer Genome Atlas (TCGA) and LinkedOmics portals.

\subsection*{Competing interests}
The authors declare that they have no competing interests.

\subsection*{Funding}
Not applicable.

\subsection*{Authors' contributions}
AH developed the research framework, designed the methodology, conducted the experiments, analysed the results, and drafted the manuscript. AA, MP, and AB provided supervision, methodological guidance, critical feedback, and contributed to the review and revision of the manuscript. All authors read and approved the final manuscript.

\subsection*{Acknowledgements}
Not applicable.
\bibliography{sn-bibliography} 
\end{document}